\documentclass[useAMS,usenatbib]{mn2e}
\usepackage[psamsfonts]{amssymb}
\usepackage[dvips]{graphicx}

\def\etal{{\it et al.}}
\def\eg{{\it e.g.}}

\def\apj{{\it ApJ}}
\def\mnras{{\it MNRAS}}

\def\aap{{\it A\&A}}
\def\nat{{\it nature}}
\def\chjaa{{\it ChJAA}}

\title[The Very Early Gamma-ray Burst Afterglows Powered by Structured Jets]
{The Very Early Gamma-ray Burst Afterglows Powered by Structured
Jets}
\author[T. Yan, D. M. Wei and Y. Z. Fan]{T. Yan$^{1,2}$\thanks{E-mail:
tyan@pmo.ac.cn}, D. M. Wei$^{1,2}$ and Y. Z. Fan$^{1,2}$\\
$^1$Purple Mountain Observatory, Chinese Academy of Sciences, Nanjing 210008, China\\
$^2$National Astronomical Observatories, Chinese Academy of
Sciences, Beijing, 100012, China}
\begin{document}

\date{Received... accepted...}

\maketitle

\label{firstpage}

\begin{abstract}
If X-ray flashes (XRFs) and X-ray rich Gamma-ray Bursts(XRRGs)
have the same origin with Gamma-ray Bursts (GRBs) but are viewed
from larger angles of structured jets, their early afterglows may
differ from those of GRBs. When the ultra-relativistic outflow
interact with the surrounding medium, there are two shocks formed,
one is a forward shock, the other is a reverse shock. In this
paper we calculate numerically the early afterglow powered by
uniform jet, Gaussian jet and power-law jet in the forward-reverse
shock scenario. A set of differential equations are used to govern
the dynamical evolution and synchrotron self-Compton effect has
been taken into account to calculate the emission. In uniform
jets, the very early afterglows of XRRGs and XRFs are
significantly lower than GRBs and the observed peak times of RS
emission are longer in interstellar medium environment. The RS
components in XRRGs and XRFs are difficult to be detected. But in
stellar wind, the reduce of very early flux and the delay of RS
peak time are not so remarkable. In nonuniform jet(Gaussian jet
and power-law jet), where there are emission materials on the line
of sight, the very early light curve resembles
isotropic-equivalent ejecta in general although the RS flux decay
index shows notable deviation if the RS is relativistic(in stellar
wind).

\end{abstract}


\section{Introduction}

Late time Gamma-ray Burst (GRB) afterglow emission has been well
observed and studied since the first detection in
1997\citep{p97nature, c97nature}. In contrast, afterglow shortly
after or during the main GRB is still quite uncertain because of
the lack of observation. This early afterglow may provide
important information about initial parameters of the burst and
shed light on explosion mechanism. SWIFT is able to observe
multi-wavelength radiation rapidly after $\gamma$ ray triggering.
With new observations, the early stage should be investigated more
profoundly. In the standard baryonic GRB fireball model, prompt
GRB emission is produced by collisions of internal shocks while
afterglow comes from interactions between burst-ejected materials
and circum-burst medium. After the internal shock phase, as the
fireball is decelerated by the circumburst medium, usually a pair
of shocks develop \citep{mr97apj, sp99apj}. One is a forward shock
(FS) expands outward to heat the external medium, the other is a
reverse shock (RS) propagates into the ejecta and heats this cold
initial shell. The central energy is gradually transmitted to
outer medium through shocked mediums and shocked ejecta. The early
afterglow studied in this paper is during this transition stage.

\citet{mr97apj} pointed out RS should emit detectable optical
synchrotron photons. This prediction has been confirmed by the
optical flash and the radio flare detected in GRB
990123\citep{a99nature, k99apj, sp99apj, mr99mnras, np04mn}. The
later more detailed investigation suggests that the RS region may
be magnetized \citep{f02chjaa, zkm03apj,pk04mnras}. The RS
emission of GRBs in stellar wind medium has been discussed
\citep{cl00apj, wdhl03mnras, zwd05}. The RS emission powered by
the magnetized outflow or neutron-fed outflow has been
investigated in some detail \citep{fww04aa, zk05apj, fzw05apj}.
The pair-rich reverse shock emission has been discussed by
\citet{liz03apj} and \citet{mkp05mnras}. Unfortunately, up to now,
there are only a few additional candidates. They are GRB 021211
\citep{f03apj, liw03apjl, w03aa, kp03mnras}, GRB 041219a
\citep{b05nature, fzw05apjl} and GRB 050525a \citep{k05, sd05}.
Whether the lack of RS emission events is due to observational
limits or theoretical problems may be settled by further
observation.

X-ray Flash (XRF) is an interesting phenomenon which resembles
GRB. It is similar to GRB in many aspects except its lower peak
energy ($\sim$10keV) and lower isotropic energy
($\sim10^{51}$ergs). Several models are proposed\citep{dcb99apj,
h01, hdl02mnras, b03aa, ldg05apj} among which the structured jet
model is widely discussed. When a jet is ejected and propagating,
observers on different viewing angles will see different
phenomena. GRBs are observed near the jet center. XRFs may be
detected at the edge of the jet. Between them are the X-ray rich
GRBs (XRRGs) whose peak energies and isotropic energies are
between those of GRBs and XRFs. Possible jet models for the XRFs
and XRRGs include the off-beam uniform jet model \citep{in01apj,
yin02apj}, the Gaussian jet model \citep{zm02apj, ldz04apj,
zdl04apj}, and the power-law jet model\citep{mrw98apj, jw04chjaa}.
The very early afterglow powered by structured jets has been
calculated by \citet{fww04mnras} analytically. In this work, we
present our numerical results.

The dynamical evolution is described in Section 2. The radiation
calculation procedure is shown in Section 3. Afterglows from jets
are prsented in Section 4. In section 5, our results are summarized with
some discussions.

\section{dynamical evolution}

Consider a homogenous cold shell ejected from central engine with
isotropic energy $E_{iso}$ and width $\Delta_0$. It expands
relativistically with Lorenz factor $\eta$ and contains mass
$m_{ej}=E_{iso}/\eta c^2$. When a FS and a RS are formed, there is a shocked region consists of tow parts:
the shocked medium and the shocked shell separated by the contact
discontinuity surface. The shocked region is assumed to have
homogenous bulk velocity and energy density, that is,
$\gamma_2=\gamma_3$, $e_2=e_3$ (hereafter the subscript 2 denotes
the shocked medium, 3 denotes the shocked shell and 4 denotes the
cold ejecta). However, the number density and random particle
velocity in the comoving frame are discontinuous at the contact
discontinuous surface. The random Lorentz factor of the shocked protons are
$(\gamma_2-1)\approx\gamma_2$ and $(\gamma_{34}-1)$ in regions 2 and 3, respectively
\citep{bm76}, where $\gamma_{34}$ is the
relative Lorentz factor of the shocked shell to the cold shell.

Kinetic energies of the shocked medium, the shocked shell and the
unshocked shell are respectively: {\setlength\arraycolsep{2pt}
\begin{eqnarray*}
E_2&=&(\gamma_2-1)m_2c^2+(1-\epsilon_2)\gamma_2U_2,\\
E_3&=&(\gamma_3-1)m_3c^2+(1-\epsilon_3)\gamma_3U_3,\\
E_4&=&(\eta-1)(m_{ej}-m_3)c^2,
\end{eqnarray*}}where $U_2=(\gamma_2-1)m_2c^2$ and $U_3=(\gamma_{34}-1)m_3c^2$ are
internal energies in the comoving frame. If a fraction $\epsilon$
of the fresh shocked material energy is radiated, the radiated
thermal energies of the shocked medium and shocked shell are
$\epsilon_2\gamma_2(\gamma_2-1)dm_2c^2$ and
$\epsilon_3\gamma_3(\gamma_{34}-1)dm_3c^2$. With energy
conservation, we have {\setlength\arraycolsep{2pt}
\begin{eqnarray}
d(E_2+E_3+E_4)&=&-\epsilon_2\gamma_2(\gamma_2-1)dm_2c^2 \nonumber\\
 & &-\epsilon_3\gamma_3(\gamma_{34}-1)dm_3c^2.\label{eq:con}
\end{eqnarray}}

If GRBs are located in interstellar medium(ISM), the number
density of the external medium $n_1$ is constant. If GRBs are born
in stellar wind, $n_1=AR^{-2}$, where $A=\dot{M}/4\pi
m_pv_w=3\times10^{35}A_*cm^{-1}$ and
$A_*=(\dot{M}/1\times10^{-5}M_{\odot}yr^{-1}(v_w/10^3km/s)^{-1}$\citep{cl00apj}.
The swept-up mass of FS evolves as
\begin{equation}
\frac{dm_2}{dR}=4\pi R^2n_1m_p. \label{eq:mass2}
\end{equation}

The comoving number density of the ejecta
$n_4=M_{ej}/[4\pi m_pR^2\Delta\eta]$, where
$\Delta=max(\Delta_0,R/\eta^2)$ is the width of shell considering
spreading effect. When $R$ is smaller than the spreading radius
$R_s=\Delta_0\eta^2$, $\Delta\approx\Delta_0$. If $R>R_s$,
$\Delta\approx R/\eta^2$. The swept-up mass of RS
evolves as {\setlength\arraycolsep{2pt}
\begin{eqnarray}
\frac{dm_3}{dR}&=&4\pi R^2(\beta_4-\beta_{RS})\eta
\,n_4m_p,\label{eq:mass3}
\end{eqnarray}}where $\beta_{RS}=\frac{\gamma_3n_3\beta_3-\gamma_4n_4\beta_4}
{\gamma_3n_3-\gamma_4n_4}$, which is the Lorentz factor of RS \citep{fww04aa}.
From equations (\ref{eq:con}-\ref{eq:mass3}), it is found that
$\gamma_2$ evolves as
\begin{equation}\label{eq:gamma}
\frac{d\gamma_2}{dR}=-4\pi R^2 \frac{Q}{P},
\end{equation}where
{\setlength\arraycolsep{2pt}
\begin{eqnarray*}
Q&=&(\gamma_2^2-1)n_1m_p+(\gamma_2\gamma_{34}-\eta)(\eta
n_4m_p)(\beta_4-\beta_{RS}),\\
P&=&m_2+m_3+(1-\epsilon_2)(2\gamma_2-1)m_2+(1-\epsilon_3)(\gamma_{34}-1)m_3\\
 & &+(1-\epsilon_3)\gamma_2m_3[\eta(1-\beta_2\beta_4)-\frac{\eta\beta_4}{\gamma_2^2\beta_2}].
\end{eqnarray*}}The overall dynamical evolution of RS-FS
can be obtained by solving equations
(\ref{eq:mass2}-\ref{eq:gamma}) combined with
\begin{equation}\label{eq:time}
dR=\frac{\beta_2}{1-\beta_2}\frac{cdt}{1+z}
\end{equation}where $t$ is
the time measured by the observer and $z$ is the redshift.

Define $\eta_e$ to be the electron radiative efficiency, that is,
a fraction $\eta_e$ of the electron energy is radiated. In the
fast cooling stage, electrons cool immediately and $\eta_e=1$. In
the slow cooling stage, $\eta_e=(\gamma_m/\gamma_c)^{p-2}$
\citep{se01apj}, where $\gamma_m$ and $\gamma_c$ are the minimum
and the cooling Lorentz factor of the shocked electrons,
respectively (see detail in the next section). Usually it is
assumed that electrons share a fraction $\epsilon_e$ of internal
energy. Then the radiative efficiency is
\begin{equation}
\epsilon=\epsilon_e\eta_e=\epsilon_e\min
[1,(\gamma_m/\gamma_c)^{p-2}].
\end{equation}$\epsilon=0$ corresponds to the adiabatic case while $\epsilon=1$
corresponds to the fully radiative case.

\citet{sp95apj} introduced
$\xi\equiv(l/\Delta_0)^{1/2}\eta^{-4/3}$ to describe the strength
of the RS, where $l\equiv[3E_{iso}/(4\pi n_1m_pc^2)]^{1/3}$ is the
Sedov length. If $\xi<1$ at the RS crossing time ($t_{\times}$),
the RS is relativistic (RRS). If $\xi>1$ at $t_{\times}$, the RS
is non-relativistic (NRS). Our results are consist with them. The
scaling laws are exactly the same with analytical results though
differences exist in absolute values. In our numerical results, it
is found out that $\gamma_{34}-1$ can be well fitted by
{\setlength\arraycolsep{2pt}
\begin{eqnarray}
\lg(\gamma_{34}-1)&=&-0.64991-1.64406\lg\xi\nonumber\\
& &-0.59494(\lg\xi)^2-0.13656(\lg\xi)^3.
\end{eqnarray}}for $0.01<\xi<10$.

After the RS crosses the shell, the shocked medium and
the shocked ejecta evolve somewhat independently. The FS continues to spread outwards, collect external medium and
the shocked medium is gradually shaped into BM self-similar
profile. The energy conservation still applies:
\begin{equation}\label{eq:mass2plus}
dE_2=-\epsilon_2\gamma_2(\gamma_2-1)dm_2c^2\qquad(t>t_{\times}).
\end{equation}Then the forward shock evolution can be solved together with
equation(\ref{eq:mass2}), (\ref{eq:time}) and (\ref{eq:mass2plus}).
At the same time, the shocked ejecta ceases to increase,
\emph{i.e.}, $m_3=m_{ej}=const$ for $t>t_{\times}$. \citet{ks00apj}
showed that its evolution consists approximately with BM solution,
that is, $\gamma_3\propto R^{-7/2+k}$($k=0$ applies to ISM and
$k=2$ to wind). They also inferred that the shocked ejecta spreads
with the speed of light in the comoving frame and evolves
adiabatically.

\section{radiation}
Both FS and RS heat the cold materials to higher temperatures,
accelerate protons and electrons as well as generate random
magnetic field. Here we calculate the synchrotron radiation and
the synchrotron self-Compton cooling of the shocked electrons. The
internal energy density of FS shocked medium
$e_2=4\gamma_2^2n_1m_pc^2$\citep{bm76}. For the ejecta shocked by
RS, $e_3=e_2$ before $t_{\times}$, otherwise $e_3\propto
t^{-\frac{4(3+g)}{3(1+2g)}}$($g=7/2-k$). $\epsilon_e$ and
$\epsilon_B$ are defined to be the faction of internal energy
occupied by electrons and magnetic field. It is easy to get that
magnetic field $B=\sqrt{8\pi\epsilon_Be}$. As to electrons, in the
absence of energy loss, they are assumed to be shocked to a
power-law distribution
$dN_e/d\gamma_e\propto\gamma_e^{-p}(\gamma_m<\gamma_e<\gamma_M)$
with minimum Lorentz factor
 \[
\gamma_{m,2}=\epsilon_e(\gamma_2-1)\frac{m_p(p-2)}{m_e(p-1)}+1,
\]
\[
\gamma_{m,3}=\epsilon_e(\gamma_{34}-1)\frac{m_p(p-2)}{m_e(p-1)}+1,
\]
and maximum Lorentz factor
$\gamma_M=10^8(B/1G)^{-1/2}$\citep{dl01apj}. Here $e$ and $B$ are
all measured in the coming frame. Following \citet{spn98apj}, we
introduce the cooling Lorentz factor $\gamma_c=\frac{6\pi (1+z)
m_ec}{\sigma_T\gamma B^2 t}$ to describe the synchrotron radiation
loss of the shocked electrons. Note that
$\gamma_{c,3}=\gamma_{c,2}$ before $t_{\times}$. The actual
electron distribution should be given as following (see also
\citet{fww04aa}): (i) for $\gamma_c \leq \gamma_m$, i.e., the fast
cooling case
\begin{equation}
 \frac{dN_e}{d\gamma_e} =C_1 \left \{
   \begin{array}{ll}
 \gamma_e^{-2}\,, \,\,\,\, & (\gamma_c \leq \gamma_e
                          \leq \gamma_m), \\
 \gamma_m^{p-1}\gamma_e^{-(p+1)}\,, \,\,\,\, & (\gamma_m<\gamma_e
                          \leq \gamma_M),
   \end{array}
   \right.
\end{equation}
where
\[
C_1=[{1\over \gamma_c}-{p-1\over p}{1\over
\gamma_m}-{\gamma_m^{p-1}\gamma_M^{-p}\over p}]^{-1}N_{\rm tot}\,,
\]
where $N_{\rm tot}$ is the total number of radiating electrons
involved; (ii) for $\gamma_m < \gamma_c \leq \gamma_M$, i.e., the
slow cooling case
\begin{equation}
  \frac{dN_e'}{d\gamma_e} = C_2\left \{
   \begin{array}{ll}
\gamma_e^{-p}\,, \,\,\,\, & (\gamma_m \leq
\gamma_e\leq \gamma_c), \\
\gamma_c\gamma_e^{-(p+1)}\,, \,\,\,\, & (\gamma_c<\gamma_e\leq
\gamma_M),
   \end{array}
   \right.
\end{equation}
where
\[
C_2=[{\gamma_m^{1-p}\over p-1}-{\gamma_c^{1-p}\over
p(p-1)}-{\gamma_c\gamma_M^{-p}\over p}]^{-1}N_{\rm tot}.
\]

In the comoving frame, the synchrotron radiation of the
two-segment distributed electrons can be well described by
power-law spectrum consisted of several segments. The spectrum
peaks at $\min(\nu_m',\nu_c')$ with flux\citep{wg99apj}
\begin{equation}
F_{max}'=\Phi_p\frac{\sqrt{3}e^3BN_{\rm tot}}{m_ec^2},
\end{equation}where $\Phi_p$ is a function of p(for $p\simeq2.5, \Phi_p\simeq0.60$).
The breaking frequencies corresponding to $\gamma_m$ and
$\gamma_c$ are $\nu_m' = 3 \gamma_m^2 e B / (4 \pi m_e c)$ and
$\nu_c' = 3 \gamma_c^2 e B / (4 \pi m_e c)$. The absorption
frequency $\nu_a'$, under which the synchrotron self-absorption
can not be ignored, is calculated accordingly \citet{wdhl03mnras}.
These three breaking frequencies divide the spectrum into four
segments. The spectrum indices are, from low frequency to high
frequency, (i)for $\nu_a'<\nu_m'<\nu_c'$, $[2,1/3,(1-p)/2,-p/2]$;
(ii)for $\nu_a'<\nu_c'<\nu_m'$, $[2,1/3,-1/2,-p/2]$; (iii)for
$\nu_m'<\nu_a'<\nu_c'$, $[2,5/2,(1-p)/2,-p/2]$; (iv)for
$\nu_c'<\nu_a'<\nu_m'$, $[2,5/2,-1/2,-p/2]$; (v)for
$\nu_m'<\nu_c'<\nu_a'$ and $\nu_c'<\nu_m'<\nu_a'$,
$[2,5/2,5/2,-p/2]$.

We assume that the synchrotron power is radiated isotropically in
the comoving frame, $\frac{d F'(\nu ')}{d \Omega '} = \frac{F'(\nu
')}{4 \pi}$. The angular distribution of power in the observer's
frame is\citep{rl76, hgdl00apj}
\begin{equation}
\frac{d F(\nu)}{d \Omega} = \frac{1+z}{\gamma^3 (1 - \beta \mu)^3}
                \frac{dF'(\nu ')}{d \Omega '}
              = \frac{1+z}{\gamma^3 (1 - \beta \mu)^3}
                \frac{F'(\nu ')}{4 \pi},
\end{equation}where
\begin{equation}
\nu = \frac{\nu'}{(1+z)\gamma (1 - \mu \beta)}.
\end{equation}Then the observed flux
density at frequency $\nu$ is{\setlength\arraycolsep{2pt}
\begin{eqnarray}
S_{\nu}& = & \frac{1}{A} \left( \frac{dF(\nu)}{d \Omega}
\frac{A}{D_{\rm L}^2} \right)\nonumber\\
  &  = &\frac{1+z}{\gamma^3 (1 - \beta \mu)^3} \frac{1}{4 \pi D_{\rm L}^2}
          F'\left((1+z)\gamma(1 - \mu \beta) \nu \right),
\end{eqnarray}}where $A$ is the area of our detector and $D_{\rm L}$ is the
luminosity distance (we assume $H_0=65 \rm km$  $\rm s^{-1}$ $\rm
Mpc^{-1}$, $\Omega_M=0.3$, $\Omega_\wedge=0.7$).

After the RS crosses the shell, no new electrons are accelerated
in the shocked shell anymore and all the electrons cool at the
same rate caused by the adiabatic expansion of the fluid. No
electrons exist above $\gamma_c$, so the flux above $\nu_c'$ drops
exponentially.

At a particular time $t$, the photons received by the observer
comes from an equal arriving time surface determined by
\citep{hgdl00apj}
\begin{equation}\label{eq:eats}
t=(1+z)\int\frac{1-\beta\cos{\Theta}}{\beta c}dR\equiv
\textrm{const.},
\end{equation}
where $\Theta$ is the angle between the radiation region and the
viewing line in the burst frame.

We also consider synchrotron self-Compton (SSC) effect. It will reduce
the cooling Lorentz factor to
\begin{equation}\label{eq:cooling}
\gamma_c=\frac{\gamma_c^{syn}}{1+Y},
\end{equation}
where Y is the Compton parameter, expressed as \citep{se01apj, wdl01apj}
\begin{equation}
Y=\frac{L_{ICS}}{L_{syn}}=\frac{\eta_e\epsilon_e/\epsilon_B}{1+Y}=\frac{-1+\sqrt{1+4\eta_e\epsilon_e/\epsilon_B}}{2}.
\end{equation}
Take it into equation(\ref{eq:cooling}), $\gamma_c$ can be solved
numerically. At early time, the forward shocked material is in the
fast cooling phase and the cooling frequency will be reduced
significantly due to the SSC radiation. As to the RS, whether it
is in the fast or slow cooling phase is parameters dependent.
However, even if it is in slow cooling phase, the cooling
frequency will be suppressed by SSC if
$\epsilon_e/\epsilon_B\gg1$, which is the ordinary case.

\section{numerical results}

In this section the early afterglow from isotropic ejecta is
presented at first. Then we calculate afterglows powered by
different jets. Three kinds of jets are considered. They are the
uniform jet, the Gaussian jet, and the power-law jet. The sideways
expansion of these jets is ignored in our calculation.

The obstacle we encountered in the current work is the
poorly known initial Lorentz factor of these jets, especially for
the structured ones (The Gaussian jet and power-law jet). By now,
the spectrum of most XRFs are non-thermal. The emitting region should be optical
thin, i.e., the optical thickness $\tau=n_{\rm e}\sigma_{\rm
T}R'<1$, where $R'$ is the comoving width of the emitting region,
$n_{\rm e}\approx \frac{L} {16\pi\epsilon_{\rm e}\eta^5 m_{\rm
p}c^4\delta t R'}$ is the comoving number density of electrons
contained in the emitting region. $L$ is the isotropic luminosity
of the burst, $\delta t$ is the typical
variability timescale of the burst lightcurve. Assuming that XRFs
are powered by internal shocks, the optical
thin ($\tau<1$) condition yields (e.g., \citet{fw05})
\begin{equation}
\eta\geq 30 L_{49}^{1/5}\delta t_{-2}^{-1/5}.
\end{equation}

\citet{kg03apj} have performed a numerical investigation on the
hydrodynamical evolution of a Gaussian jet by assuming
$\eta(\theta)$ is also Gaussian. In the Monte Carlo simulation of
Zhang et al. (2004), $\eta(\theta)$ has been taken as a free
function but to find that a flat or only small fluctuation angular
distribution of Lorentz factor is indeed required. Therefore, and
partly for convenience, we take $\eta=300$ for off-axis uniform
jet, the Gaussian jet, and the power-law jet.

For each jet model, light curves with typical parameters from
three viewing angles are presented, which may correspond to GRBs,
XRRGs and XRFs respectively. The light curves differ from each
other although there are some universal properties. Viewing from
the center of a jet, the light curve is approximately the same as
that powered by an isotropic ejecta at the early time and a break
occurs when the emission from the edge comes into view. When we
observe off-axis, the flux is lower at the early time but
gradually converge into the light curve from the center when the
whole jet can be seen.

\subsection{Isotropic ejecta}
The upper panel of Fig.1 shows the early R band light curve from
isotropic ejecta in ISM case with typical parameters. The RS
dominates the very early R band light curve, peaks at about 14mag.
The RS is non-relativistic at $t_{\times}$ with
$\nu_m^{RS}<\nu_a^{RS}<\nu_R<\nu_c^{RS}$. Therefore
$t_{\times}\approx45s$ is greater than the prompt duration
$T_{dur}=(1+z){\Delta_0\over c}\approx20s$. The R band arising
index is $\sim2.5$, which is shallower than the analytical
estimation of NRS \citep{k00apj, f02chjaa}. It varies with time as
well as with initial parameters \citep{np04mn}. After
$t_{\times}$, the decay index is $\sim2$ until $\nu_c^{RS}=\nu_R$.
Then a break occurs followed by faster decay. The flux after the
break comes from equal arriving time surface emitted at lower
radius. Since $\nu_c$ is reduced by SSC, it reaches the R band and
the RS flux encounters faster decay shortly after $t_{\times}$.
Due to the rapid decay of RS emission, the later emission is
mainly contributed by FS. It is interesting to note that the late
FS emission bump vanishes when the SSC effect has been taken into
account. The physical reason is that with SSC effect, the FS flux
peak time appears earlier at $\nu_R=\nu_c^{FS}$ during fast
cooling. Since $\nu_c^{FS}\propto t^{-1/2}, \nu_m^{FS}\propto
t^{-3/2}$ and $F_{max}^{FS}\sim const$, the decay index following
the peak flux is quite flat(-1/4) and change to $-(3p-2)/4$ after
$\nu_R=\nu_m^{FS}$.

In stellar wind environment, the RS also dominates the very early
R band light curve with typical parameters, which is shown in the
lower panel of Fig.1. The peak flux is about 13mag. The RS is
relativistic at $t_{\times}$ with
$\nu_c^{RS}<\nu_a^{RS}<\nu_R<\nu_m^{RS}$. $t_{\times}\approx40s$
coincides with $T_{dur}\approx65s$. The R band arising index is
$\sim0.5$, which is much flatter than that in the ISM case. After
$t_{\times}$, the flux decay with -2.5, which is contributed to
emission from lower radius on equal arrival time surface. Since
both RS and FS are in the fast cooling case and $\max\{
\nu_c^{RS},\nu_c^{FS} \}<\nu_R$, the R band flux is reduced by a
factor $1+Y\simeq1+\sqrt{\epsilon_e/\epsilon_B}$(about 2 magnitude
taking $\epsilon_e/\epsilon_B=30$) when the SSC effect has been
taken into account.

\begin{figure}
\includegraphics[clip,width=0.95\linewidth]{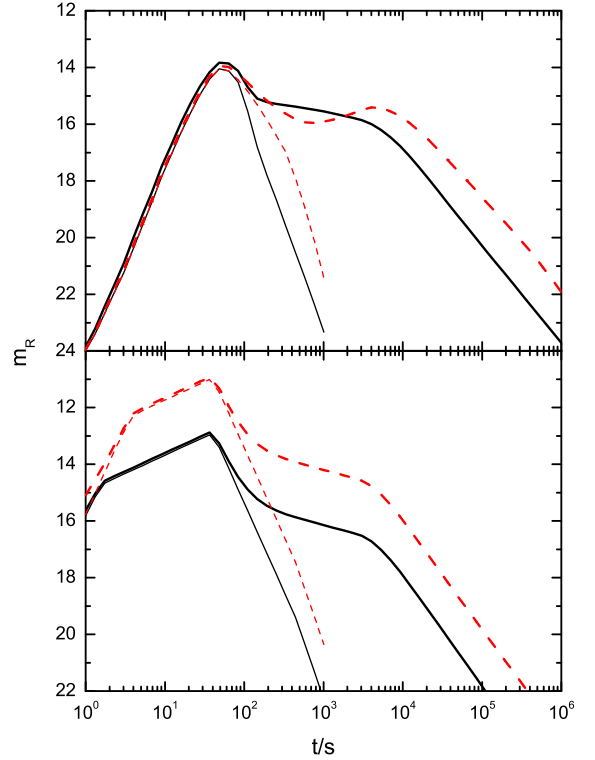}
\caption{R band afterglow from isotropic ejecta with typical
parameters. Upper panel is the ISM case and lower is the wind
case. Thin lines are the RS emission and thick lines are the total
emission. SSC effect is ignored in the dashed lines and taken into
account in the solid lines. In the ISM case, the shape of the
light curve changed since $\nu_c^{RS},\nu_c^{FS}\sim\nu_R$ around
$t_\times$. The RS shows a more rapid decay and the bump caused by
the FS vanishes. In the wind case, both RS and FS flux are reduced
since $\max(\nu_c^{RS},\nu_c^{FS})<\nu_R$. Parameters: $\eta=300$,
$\Delta_0=3\times10^{11}cm$, $E_{iso}=1\times10^{53}ergs$,
$n_1=1cm^{-3}$ for ISM and $\eta=300$,
$\Delta_0=1\times10^{12}cm$, $E_{iso}=5\times10^{52}ergs$,
$A_*=1.0$ for wind, $\epsilon_e=0.3, \epsilon_B=0.01, p=2.5,
z=1$.} \label{ics}
\end{figure}

\subsection{Uniform jet}
\begin{figure}
\includegraphics[clip,width=0.95\linewidth]{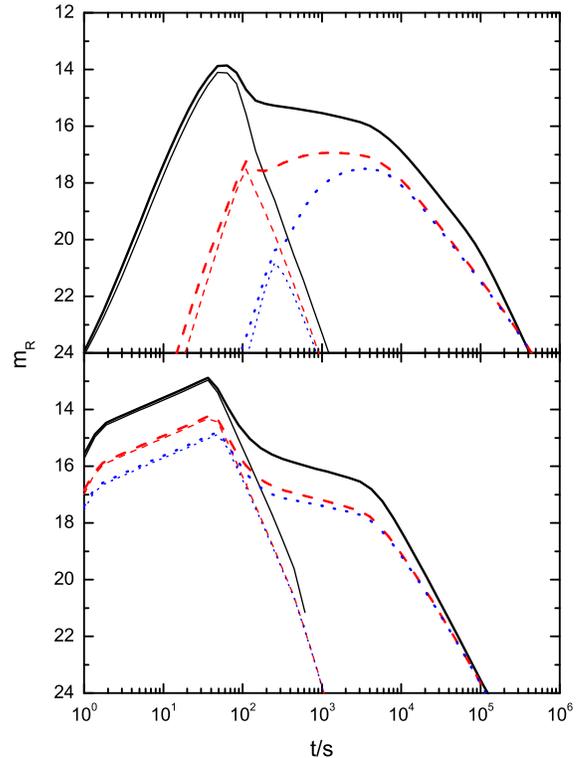}
\caption{R band afterglow from uniform jet. Upper panel is the ISM
case and lower is the wind case. Thin lines are the RS emission
and thick lines are the total emission. $\Delta
\Theta=0,~3/2\eta,~3/\eta$ in solid lines, dashed lines and dotted
lines respectively. Parameters: $\theta_{jet}=0.1$, other
parameters are the same as in Fig.1. } \label{uniform}
\end{figure}

Uniform jet is a simple model which assumes that all the ejected
materials are confined to a collimated uniform cone. GRBs are
detected within the cone (on-beam). XRFs and XRRGs are viewed out
of the jet(off beam). If our viewing angle $\Theta_v$ is slightly
larger than the jet opening angle $\theta_{jet}$, the observed
frequency
\begin{equation}
\nu_{off}=a\nu_{on},
\end{equation}where $a\approx[1+\gamma^2(\Delta\Theta)^2]^{-1}$
($\gamma$ is the bulk Lorentz factor of the jet) and $
\Delta\Theta=\Theta_v-\theta_{jet}$. Taking the peak energy of
XRFs $E_{p,XRF}\approx 0.1E_{p,GRB}$, XRFs should be observed at
$\Delta\Theta_{XRF}\approx3/\eta$. The observed flux can be
estimated by an empirical formula (e.g., \citet{fww04mnras})
\begin{equation}
F_{\nu_{off}}(\Delta\Theta,t_{obs})\approx
{a^3\over2}F_{\nu_{on}}(0,t),
\end{equation}where $dt_{obs}=dt/a$. It implies that if
$\gamma>1/\Delta\Theta$,the observed flux is dramatically
lower and the observed time is longer. The early afterglow powered by the
uniform jets has been presented in Fig. \ref{uniform}.

In the ISM case, the flux increases rapidly at the early time. The
RS component reaches its peak flux at a time longer than viewing
on-beam. The peak flux of RS emission for $\Delta \Theta
=3/(2\eta)$ (i.e., the XRRG) is about 17.5mag at about 100s. It is
bright enough to be detected. The rising index of RS is about 3.5.
The peak flux of the RS emission for  $\Delta \Theta =3/\eta$
(i.e., the XRF) is about 21mag at about 250s. It is hard to be
detected by UVOT on SWIFT or ground telescope at such early time.
The rising index is about 4.2. After $t_{\times}$, the observed FS
flux keeps rising for some while, peaks several hours later. The
peak FS emission flux of XRRG is about 17mag and 17.5mag for XRF.
Therefore in XRFs, the RS component can hardly be seen and the
later FS component around 1hour is much more luminous. In XRRG,
the RS emission is detectable but the RS component is not
significant. The light curve is nearly flat for $t<$1 hour.

In the wind case, the off-beam RS very early flux increases at the
same rate as in the on-beam case since both $\gamma$ and $a$ are
constant, but the flux is lower. The peak time of off-beam RS is
slightly longer, $\sim50s$ in XRF. The XRF peak flux is about
15mag and significantly higher than the FS. After $t_{\times}$, as
the Lorentz factor decreases, the light curve gradually converge
into the light curve from the center.

\subsection{Nonuniform jet}
Different from uniform jets, there are emission materials on the
line of sight in nonuniform jets(Gaussian jets and power-law
jets). XRFs and XRRGs are viewed within the jets but have lower
isotropic-equivalent energy than GRBs. The observed early emission
is mainly contributed from materials on the line of sight.
However, we find that the light curve show notable deviation if
the RS is relativistic.

\subsubsection{Gaussian jet} The jet energy in a Gaussian jet
distributed with angular as
\begin{equation}
E(\theta)=E_0e^{-\theta^2/2\theta_0^2}{\rm
~~~~~}0\leq\theta\leq\theta_{jet}.
\end{equation}
With typical value that $E_{iso,XRF}=10^{-2}E_{iso,GRB}$, the
viewing angle of XRF should be
$\Theta_{v,XRF}\approx3.04\theta_0$. We also calculate the light
curve with $\Theta_{v,XRRG}\approx2.15\theta_0$ which may
correspond to the XRRG(taking $E_{iso,XRRG}=10^{-1}E_{iso,GRB}$).
The R band afterglow from Gaussian jet is shown in Fig.3.

In the ISM case, the off-axis RS flux is exactly the same with
isotropic emission taking $E_{iso}=E(\Theta_v)$. The RS is
non-relativistic and the Lorentz factor is high at early time. So
the jet structure can not change the light curve significantly. As
$\Theta_v$ increases, the light curve has (i) shorter
$t_{\times}$; (ii)lower peak RS emission flux, about 17mag for
XRRG and 19.5mag for XRF; (iii) flatter rising before
$t_{\times}$; (iv) longer $t_{Rc}$(the time $\nu_R=\nu_c$),
therefore longer time interval for the $\sim-2$ delay after
$t_{\times}$.

In stellar wind, the off-axis RS flux is slightly different with
the isotropic emission taking $E_{iso}=E(\Theta_v)$. The RS is
relativistic and the Lorentz factor is sufficiently low to reflect
the jet structure. The isotropic emission with decreasing
$E_{iso}$ from GRB to XRF has (i)nearly constant $t_{\times}$,
constant rising index 0.5 before $t_{\times}$ and decay index -2.5
after $t_{\times}$; (ii)lower peak flux, about 16.5mag for XRRG
and 19mag for XRF. What actual light curve differs from isotropic
one is the decay index of RS flux. Viewing from the center, RS
flux decays faster. With increasing viewing angle, the start time
of the faster decay increases, and from $t_{\times}$ to this time,
the flux is flatter than isotropic emission. The surplus emission
comes from the center of the jet with $\theta<\Theta_v$.

\begin{figure}
\includegraphics[clip,width=0.95\linewidth]{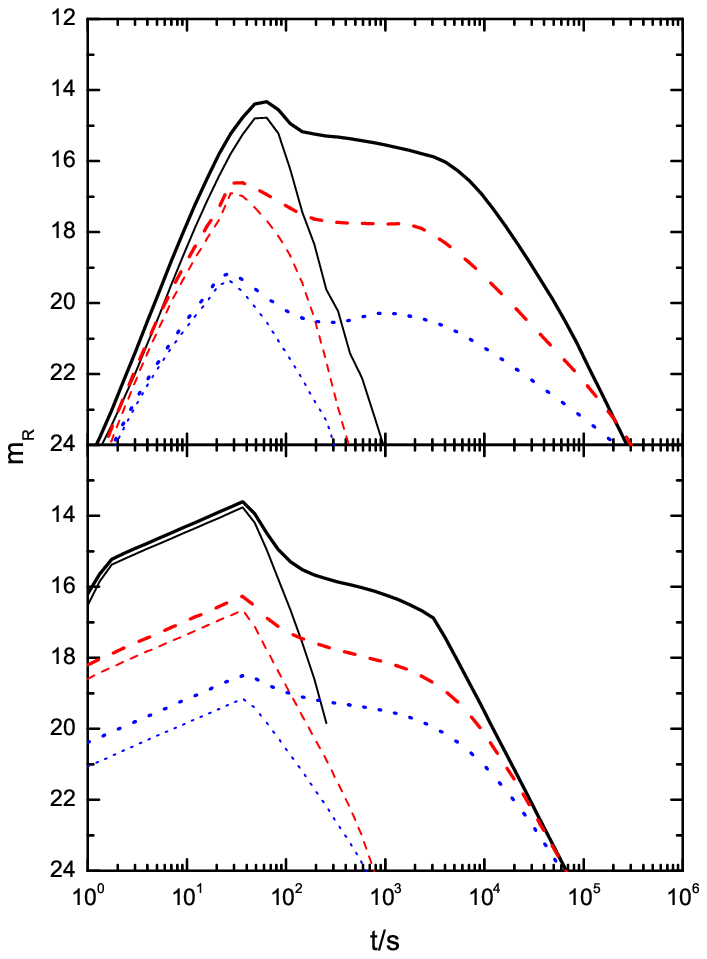}
\caption{R band afterglow from Gaussian jet. Upper panel is the
ISM case and lower is the wind case. Thin lines are the RS
emission and thick lines are the total emission.
$\Theta_v=0,2.15\theta_0,3.04\theta_0$ in solid lines, dashed
lines and dotted lines respectively. Parameters:
$E_0=1\times10^{53}ergs$ for ISM and $E_0=5\times10^{52}ergs$ for
wind, $\theta_0=0.05$,$\theta_{jet}=0.2$, other parameters are the
same as in Fig.1.} \label{Gaussian}
\end{figure}

\subsubsection{Power-law jet} In this model, the jet energy
distribution as a function of angular is as follows
\begin{equation}
E(\theta)=\left\{\begin{array}{ll}
        E_c & 0\leq\theta<\theta_c\\
        E_c(\theta/\theta_c)^{-2} &
        \theta_c\leq\theta\leq\theta_{jet}.
        \end{array}\right.
\end{equation}
$E_{iso,XRF}=10^{-2}E_{iso,GRB}$ and
$E_{iso,XRRG}=10^{-1}E_{iso,GRB}$ correspond to
$\Theta_{v,XRF}\approx10\theta_c$ and
$\Theta_{v,XRRG}\approx3.16\theta_c$ respectively. Early afterglow
from power-law jet is shown in Fig.4.

In the ISM case, the RS is also exactly the same with isotropic
emission taking $E_{iso}=E(\Theta_v)$, as found in the Gaussian
jet model. In the wind case, RS flux also resembles that in the
Gaussian jet except the decay index departure from isotropic
emission is more prominent.

\begin{figure}
\includegraphics[clip,width=0.95\linewidth]{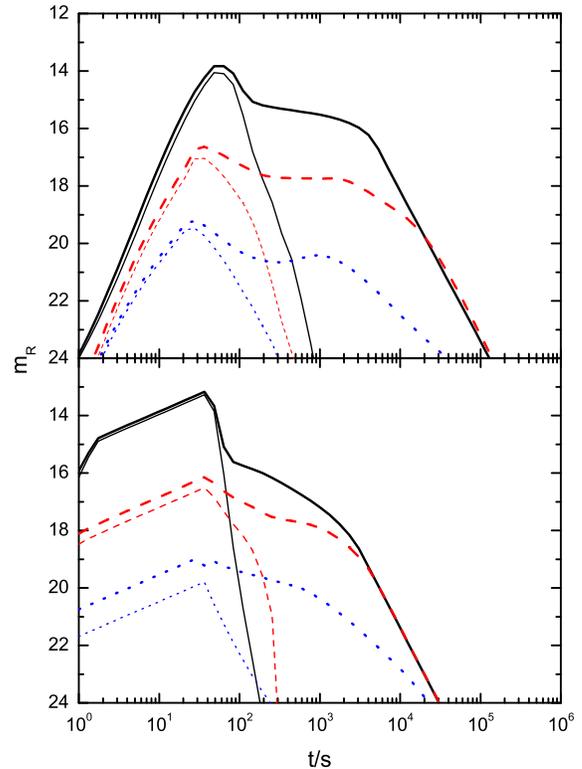}
\caption{R band afterglow from power-law jet. Upper panel is the
ISM case and lower is the wind case. Thin lines are the RS
emission and thick lines are the total emission.
$\Theta_v=0,3.16\theta_c,10\theta_c$ in solid lines, dashed lines
and dotted lines respectively. Parameters:
$E_c=1\times10^{53}ergs$ for ISM and $E_c=5\times10^{52}ergs$ for
wind, $\theta_c=0.015$, $\theta_{jet}=0.2$, other parameters are
the same as in Fig.1.} \label{power-law}
\end{figure}

\section{conclusion and discussion}
So far, there are several RS emission candidates reported in SWIFT
era. One is GRB 041219a detected by INTEGRAL. \citet{fzw05apjl}
fitted its early IR afterglow with RS-FS emission model and found
the RS region is magnetized. GRB 050525a may also be an possible
candidate. \citet{sd05} used RS-FS in standard scenario to fit the
early optical bump and the parameters are reasonable.

However, among all the bursts targeted in optical band within few
minutes after the prompt emission, only a small fraction of them
have likely RS emission. This is conflicted with the theoretical
estimation. It may be caused by the extinctions of the host
galaxy. And as demonstrated in this paper, the overestimation of
the $\nu_m$, $\nu_c$ may lead to a dimmer RS radiation. Up to now,
the very early optical afterglows of XRFs have not been well
detected, future observation may help us to modify the present
theory.

In this paper, we calculate the early afterglow powered by various
kinds of jets numerically. Dynamical evolution is solved from a
set of differential equations. This is different from the
analytical treatment of \citet{fww04mnras} and gives similar but
more exactly results. We find that the most unprecise estimation
comes from $\gamma_{34}-1$, which results in the overestimation of
$\nu_m$ and too rapid increasement of $\nu_m$ before $t_\times$ in
previous treatment. This is especially significant in the ISM case
since the RS is mildly relativistic in typical parameters taken
here. At the same time, $\nu_c$ was overestimated previously
because of the ignoring of SSC effect. Considering these two
factors, the peak flux of reverse shock should be dimmer, which
may help us to explain the lack of detection of optical flashes in
most GRBs. It also needs to be pointed out that electrons may be
cooled not only by SSC process but also by external photons( \eg,
prompt emission) through inverse Compton scattering. Then $\nu_c$
may be reduced further and lead to even fainter very early optical
radiation.

Early afterglow from jets, both uniform and structured, are
calculated. We find that the early afterglow varies significantly
with different viewing angles and is dependent on the jet
structure.

SWIFT XRT detection find that the early X-ray flare may be an
common characteristic of GRB X-ray afterglow. \citet{fw05} suggested
that the RS synchrotron emission hardly can produce the very early
X-ray flares. Our calculation confirms their results. We also
tried to simulate it with RS SSC emission within a large
parameters space but failed. The RS SSC X-ray emission is always
lower and generally far lower than FS synchrotron emission.

\section*{Acknowledgments}
We thank Zou, Y. C. and Jin, Z. P. for helpful discussion. This
work is supported by the National Natural Science Foundation
(grants 10225314 and 10233010) of China, and the National 973
Project on Fundamental Researches of China (NKBRSF G19990754).

\label{lastpage}

\end{document}